\PassOptionsToPackage{hyphens}{url}
\documentclass[
11pt,
aps,
pra,
tightenlines,
longbibliography,
showpacs,
nofootinbib,
notitlepage,
superscriptaddress]{revtex4-2}

\usepackage{amsmath}
\usepackage{amssymb}
\usepackage[dvips]{graphicx}
\usepackage{color}
\usepackage{tabularx}
\usepackage{mathtools}
\usepackage{algpseudocode}
\usepackage{enumitem}
\usepackage{multirow}
\usepackage{epstopdf}  
\usepackage{bm}
\usepackage{longtable}
\usepackage{booktabs}

\usepackage{makecell}
\usepackage[version=3]{mhchem}
\usepackage{tikz}
\usetikzlibrary{shapes,snakes}

\usetikzlibrary{arrows}
\usepackage{threeparttable}
\usepackage{hyperref}

\newcommand{\ucite}[1]{$^{\text{\!\cite{#1}}}$} 

\begin{document}

\title{Revealing the dynamic responses of Pb under shock loading based on DFT-accuracy machine learning potential}

\author{Enze Hou}
\affiliation{Institute of Applied Physics and Computational Mathematics, Beijing 100094, China}
\affiliation{Graduate School of China Academy of Engineering Physics, Beijing 100088, China}

\author{Xiaoyang Wang}
\affiliation{National Key Laboratory of Computational Physics, Institute of Applied Physics and Computational Mathematics, Fenghao East Road 2, Beijing 100094, China}

\author{Han Wang}
\email{wang\_han@iapcm.ac.cn}
\affiliation{National Key Laboratory of Computational Physics, Institute of Applied Physics and Computational Mathematics, Fenghao East Road 2, Beijing 100094, China}
\affiliation{%
HEDPS, CAPT, College of Engineering and School of Physics, Peking University, Beijing 100871, China
}

\begin{abstract}
Lead (Pb) is a typical low-melting-point ductile metal and serves as an important model material in the study of dynamic responses. 
Under shock-wave loading, its dynamic mechanical behavior comprises two key phenomena: plastic deformation and shock induced phase transitions. 
The underlying mechanisms of these processes are still poorly understood.
Revealing these mechanisms remains challenging for experimental approaches.
Non-equilibrium molecular dynamics (NEMD) simulations are an alternative theoretical tool for studying dynamic responses, as they capture atomic-scale mechanisms such as defect evolution and deformation pathways.
However, due to the limited accuracy of empirical interatomic potentials, the reliability of previous NEMD studies is questioned. 
Using our newly developed machine learning potential for Pb-Sn alloys, we revisited the microstructure evolution in response to shock loading under various shock orientations.
The results reveal that shock loading along the [001] orientation of Pb exhibits a fast, reversible, and massive phase transition and stacking fault evolution.
The behavior of Pb differs from previous studies by the absence of twinning during plastic deformation.
Loading along the [011] orientation leads to slow, irreversible plastic deformation, and a localized FCC-BCC phase transition in the Pitsch orientation relationship.
This study provides crucial theoretical insights into the dynamic mechanical response of Pb, offering a theoretical input for understanding the microstructure-performance relationship under extreme conditions.
\end{abstract}

\date{\today}

\maketitle

\section{Introduction}
As a typical group IV metal, lead (Pb) has a low ambient-pressure melting point of approximately 600~K. 
Lead has a complex pressure-temperature phase diagram, 
Pb at ambient pressure has a ductile FCC phase and transforms to the HCP phase at $\sim$13 GPa.\ucite{takahashi1969lead}
At $\sim$109 GPa, Pb transforms to the BCC phase. 
Both the FCC-HCP and HCP-BCC transitions are considered sluggish, weakly first-order transformations with minimal density changes.\ucite{KUZNETSOV2002125}
Under shock loading, the shock Hugoniot curve passes through all three phases, leading to complex dynamic behaviors, including plastic deformation\ucite{RN917} and solid-state phase transitions\ucite{RN913} at high strain rates.
These unique properties make lead a key model material for studying materials response under dynamic loading.

Experimental studies of dynamic response of Pb have focused primarily on the macroscopic dynamic fracturing. 
For example, Chen et al.\ucite{RN919} studied the mass and density distribution of fragmented Pb during its micro-spallation. 
Antipov et al.\ucite{RN921} investigated the fragment of explosive shocked Pb and quantitatively determined ejecta, spallation, and undamaged regions. 
However, due to the lack of experimental approaches that characterize the micro-structures under extreme conditions, the atomistic-scale processes under the dynamic loading conditions, such as the defect evolution, phase transition, and void nucleation, remain poorly understood. 
Non-equilibrium molecular dynamics (NEMD) simulations serve as a powerful alternative tool for investigating dynamic responses under shock loading.
For example, Xiang et al.\ucite{RN909,RN910,RN911} investigated shock-induced spallation of nano-crystal Pb, Wang et al.\ucite{wang2019atomic}, explored the spall response of release-melted lead. Mayer et al.\ucite{Mayer2020} simulated the sensitivity of spall strength to strain rates.
These simulations complement and validate the experimental findings.

Importantly, NEMD simulations can resolve the microstructure evolution during the short period of dynamic loading prior to the dynamic fracture, which is critical to a comprehensive understanding of the microstructure-performance relationship under extreme conditions.
For example, Li et al.\ucite{RN907} studied the atomistic processes of shock-induced plasticity, which were found to be operative via the slipping and deformation twinning. 
They point out that twins formed due to the mutual impediments of the slip and cross-slips may act as nucleation sites for subsequent FCC-HCP phase transitions. 
In addition, BCC structures were observed to form along the Bain-Path, and mostly returned to FCC with further relaxation in tens of picoseconds. 

It is worth noticing that most existing NEMD studies utilize empirical interatomic potential, a highly efficient formalism that employs fixed functional forms and a limited number of parameters fitted to several selected material properties. 
A well-known issue with empirical potentials is that they can hardly describe multiple phases accurately over a wide range of thermodynamic conditions.
Meanwhile, their transferability is usually insufficient for accurately simulating a wide variety of defects.
For instance, the study by Li et al. employed the embedded atom method (EAM) potential developed by Wang et al.,\ucite{RN912} which accurately reproduces Pb's Hugoniot curve but underestimates the melting points above 40 GPa. 
Moreover, in our separate test of this EAM potential, the formation energy of the deformation twinning along the (111) plane is negative, which could possibly lead to the overestimation or unphysical formation of twins during MD simulations. 
The reliability of the simulation results could therefore be compromised by the lack of accuracies of the empirical potential.
For Pb, first-principles calculations based on density functional theory (DFT) have been extensively applied to investigate its phase structures and phase diagram. 
For instance, Liu et~al. and Cui et~al.\ucite{PhysRevB.43.1795,cui2008first} computed the critical pressures for phase transitions in Pb, with the calculated FCC–HCP and HCP–BCC transition pressures showing good agreement with experiments.\ucite{takahashi1969lead,KUZNETSOV2002125} 
In addition, Cricchio et al.\ucite{PhysRevB.73.140103} used ab initio molecular dynamics (AIMD) to determine the melting curve of Pb as a function of pressure, which also aligns well with experimental data.\ucite{vohra1990static}
Furthermore, Yu et al.\ucite{PhysRevB.70.155417} calculated the surface energy, which is consistent with experimental measurements.\ucite{bombis2002absolute} 
These agreements demonstrate the high reliability of DFT in predicting the properties of Pb.
However, AIMD cannot be directly applied to simulate dynamic mechanical responses, as the computational complexity of DFT scales cubically with the number of electrons, making such simulations prohibitively expensive for large systems and long timescales.

Fortunately, recent development of the machine-learning potential (MLP)\ucite{thompson2015spectral,shapeev2016moment,behler2007generalized,bartok2010gaussian,chmiela2017machine,schutt2017schnet,smith2017ani,han2017deep,zhang2018end,wood2017quantumaccurate,Kostiuchenko_2024,Zhao_2025,Xiong_2023} has revolutionized atomic scale simulations by offering both high accuracy and computational efficiency in research fields such as perovskites \ucite{yang2022lattice}, nuclear materials \ucite{hao2023artificial}, superconductors \ucite{li2023theoretical}, high-energy-density materials \ucite{zhang2025research}, high pressure water phases\ucite{zhuang2020discriminating,qiu2023anomalous} and atmosphere of gas giant planets\ucite{chang2025h}.
MLPs have also been widely employed in non-equilibrium dynamic simulations. 
For instance, Gao et~al.\ucite{gao2024neural} investigated the ultrafast laser-induced melting of copper nanofilms using an MLP; Zeng et al.\ucite{18-20231258} examined liquid iron under extreme conditions; Liu et al.\ucite{liu2025machine} studied the spallation behavior of p-phenylene terephthalamide crystals; D'Souza et al.\ucite{10.1063/5.0254638} explored the behavior of deuterium under shock loading; Pan et al.\ucite{PhysRevB.110.224101} analyzed shock compression pathways in pyrite silica; and Zeng et al.\ucite{Zeng2025A} researched the shock-induced phase transition of solid iron at high pressures. These examples collectively demonstrate the applicability and reliability of MLPs in simulating the dynamic responses of materials. 

We recently developed a machine-learned potential (MLP) named DP-PbSn,\ucite{DP-PbSn} tailored for Pb-Sn alloy with up to 30 at.\% Sn. 
With the aid of the non-von Neumann molecular dynamics (NVNMD) architecture,\ucite{NVNMD} this potential is able to access large-scale NEMD simulations including millions of atoms with first-principles accuracy over a wide thermodynamic range (0–5000 K, 0–100 GPa). 
In this work, we employ this potential in NEMD simulations to revisit the atomistic processes in Pb's dynamic response, including the shock induced plasticity and the evolution of the BCC phase during shock loading. 
Meanwhile, the anisotropy of shock-induced plastic deformation and phase transformation of Pb is investigated. 

The findings in this study may provide new mechanistic insights into the dynamic mechanical response of Pb.
Meanwhile, the techniques used in this study can be applied to investigating the dynamic processes of other materials.

\section{Methods}
In this section, we first briefly introduce the MLP, then present the molecular dynamics setups employed in this study.
\subsection{The MLP: DP-PbSn}
The MLP employed in this study, namely DP-PbSn, was developed based on the DeepMD-Kit\ucite{wang2018deepmd,zhang2018end} framework. 
This potential was trained on a dataset comprising over 8217 configurations, covering a thermodynamic range of 0–100 GPa and 0–5000 K.
The dataset encompassed molecular dynamics trajectories of BCC, FCC, HCP, and liquid phases of Pb-Sn alloy, sampled both during the evolution toward equilibrium and after the equilibrium under the target thermodynamic conditions. These phases are experienced by the material along the shock Hugoniot during compression.
Moreover, the structures of mono-vacancy and free surfaces were included in the training dataset, which enhances the reliability of the potential in describing fracture nucleation.
For dataset generation, the DFT calculations utilized a plane-wave basis set implemented in the ABACUS code.\ucite{abacus1,abacus2} 
The Perdew-Burke-Ernzerhof (PBE) exchange-correlation functional within the generalized gradient approximation (GGA)\ucite{Perdew1996PBE} was used.     
An energy cutoff of 100~Ry and a Monkhorst-Pack \textit{k}-point sampling grid spacing of 0.15~Å\(^{-1}\) were selected.
The Optimized Norm-Conserving Vanderbilt (ONCV) pseudopotentials\ucite{oncv}  from PseudoDojo\ucite{PD04,PDTh} were employed.

The MLP achieved a testing RMSE on an independent test set of 2.27 meV/atom, 40.6 meV/Å, and 25.8 meV/atom for energy, force, and virial tensor, respectively. The detailed DP-PbSn characterization results are presented in Ref.\cite{DP-PbSn} 

\begin{center}
\includegraphics[width=15cm]{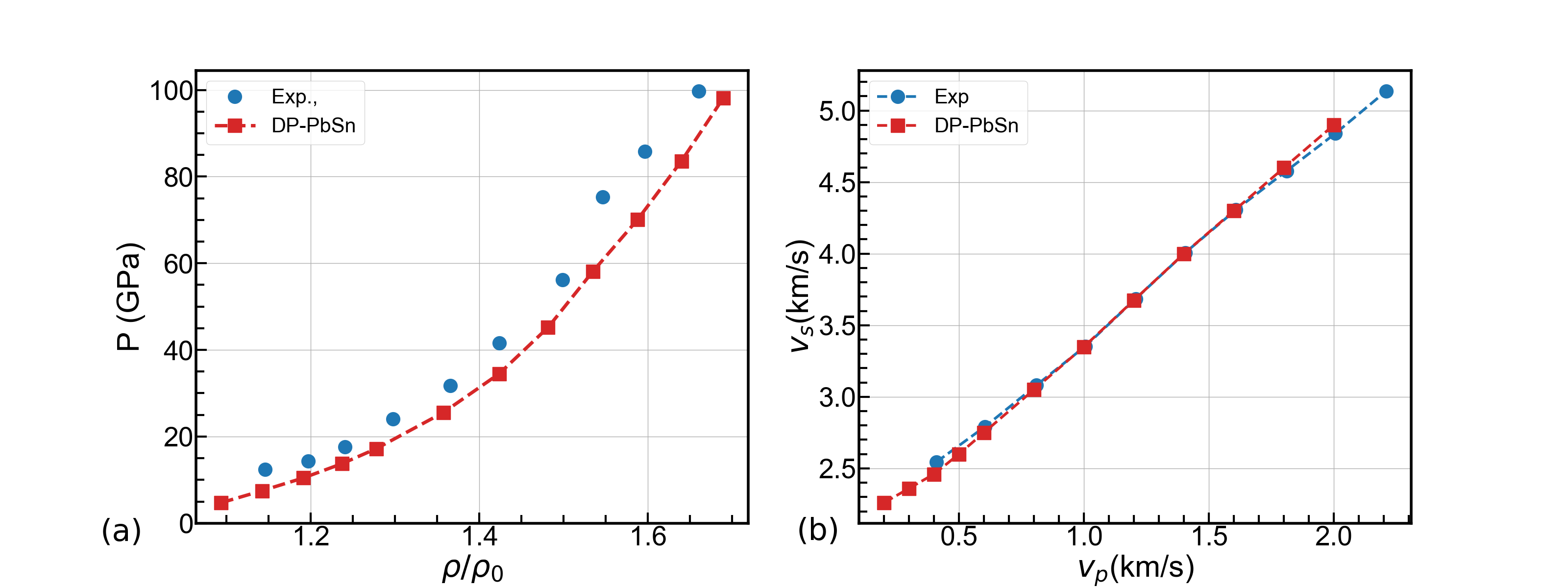}\\[8pt]
\parbox[c]{15.0cm}{\footnotesize{\bf Fig.~1.} 
The Hugoniot curve predicted by DP-PbSn. 
(a) the pressure-density relationship compared with the experimental result from Ref.\cite{marsh1980lasl}. 
(b) The shockwave velocity-piston velocity relationship compared with the experimental results.\ucite{marsh1980lasl} 
These results were reproduced from Ref.\cite{DP-PbSn} with the permission of the authors. }
\end{center}

The Hugoniot curve of Pb calculated using NEMD simulations is presented in Fig.~1. 
While the pressure-density (P-$\rho$) relationship along the Hugoniot predicted by DP-PbSn followed the general trend of the experimental data (Fig.~1 (a)), it exhibited a nearly constant shift towards a lower pressure. 
The observed deviation between the calculated Hugoniot curve and experimental data likely stems from systematic errors in DFT calculations when compared to experimental results. 
As evidenced by the equation of state reported in Ref.\cite{wang2019atomic}, DFT calculations may consistently underestimate the pressure relative to experimental measurements at equivalent volumes. 
This systematic underestimation aligned with the discrepancies shown in our results.
The particle velocity-wave velocity ($v_\mathrm{p}-v_\mathrm{s}$) relationship obtained with DP-PbSn is shown in  Fig.~1 (b), demonstrating near-linear behavior that closely matched the experimental measurements.\ucite{marsh1980lasl} 
These Hugoniot curve results further validate the applicability of DP-PbSn for investigating dynamic mechanical responses.

\subsection{Molecular Dynamics Setup}
We performed NEMD simulations on FCC-Pb samples with dimensions of approximately 100~Å × 100~Å × 1,000~Å along the x, y, and z axes. 
The samples comprised 320,000 atoms. 
The simulation cell size was sufficient to capture the dynamic processes from shock compression to the onset of dynamic fracture, while also representing a practical compromise given the substantial computational cost associated with the MLP.
Prior to dynamic loading, the sample was equilibrated at 300 K and zero pressure with the $NPT$ ensemble for 50 ps. 
Periodic boundary conditions were applied along all dimensions during the equilibrium simulation.

In the non-equilibrium simulation, dynamic loading was applied via a piston at the lower z boundary of the sample, with a controlled particle velocity $v\mathrm{p}$ ranging from 0.2 to 1.4~km/s, generating a compressive shock wave that propagated toward the upper z boundary.\ucite{LWH_SiC}
An $NVE$ ensemble was applied to the system throughout the NEMD simulations. 
Periodic boundary conditions were maintained in the x and y dimensions, while the boundary condition along the z dimension was set to free.
The piston was removed immediately when the shock wave reached the upper z-boundary.
As a result, a release wave from the lower z-boundary and a rarefaction wave reflected from the upper z-boundary were initiated simultaneously.
During the NEMD simulations, we implemented an adaptive time step ranging from 0.001 fs to 1 fs, ensuring the maximum atomic displacement within any single time step did not exceed 0.1 Å.
The use of an adaptive time step effectively balanced the temporal resolution required to capture rapidly evolving dynamical processes against the computational cost associated with MLPs.

The equilibrium and NEMD simulations were implemented using the LAMMPS package.\ucite{plimpton1995lammps} 
The material properties were evaluated by analyzing thin slices along the z-direction of the system. 
The width of each slice was 2.0~Å.  
Within each slice, we computed the per-atom velocity ($v_{i}$), pressure, shear stress, and local temperature.
The shear stress was calculated as: 
\begin{equation}
    \tau = \frac{1}{2} ( \sigma_\mathrm{zz} - \frac{1}{2} \left( \sigma_\mathrm{xx} + \sigma_\mathrm{yy} \right)),
\end{equation}
where $\sigma$ is the stress tensor for each atom.
The local temperature was calculated as:
\begin{equation}
T = \frac{1}{3 N k_\mathrm{b}} \, \sum_{i=1}^{N} m_{i} \Bigl( v_{i,\mathrm{x}}^{2} + v_{i,\mathrm{y}}^{2} + (v_{i,\mathrm{z}}-v_\mathrm{com,z})^{2} \Bigr),
\end{equation}
where $v_\mathrm{com,z}$ is the velocity of the center of mass of each slice along the z-direction, which was properly reduced for temperature evaluation.

The crystal structure in the NEMD snapshots was primarily determined via the polyhedral template matching (PTM) analysis.\ucite{Larsen_2016} 
The visualization of the snapshots was performed using the OVITO software.\ucite{ovito}
\begin{center}
\includegraphics[width=15cm]{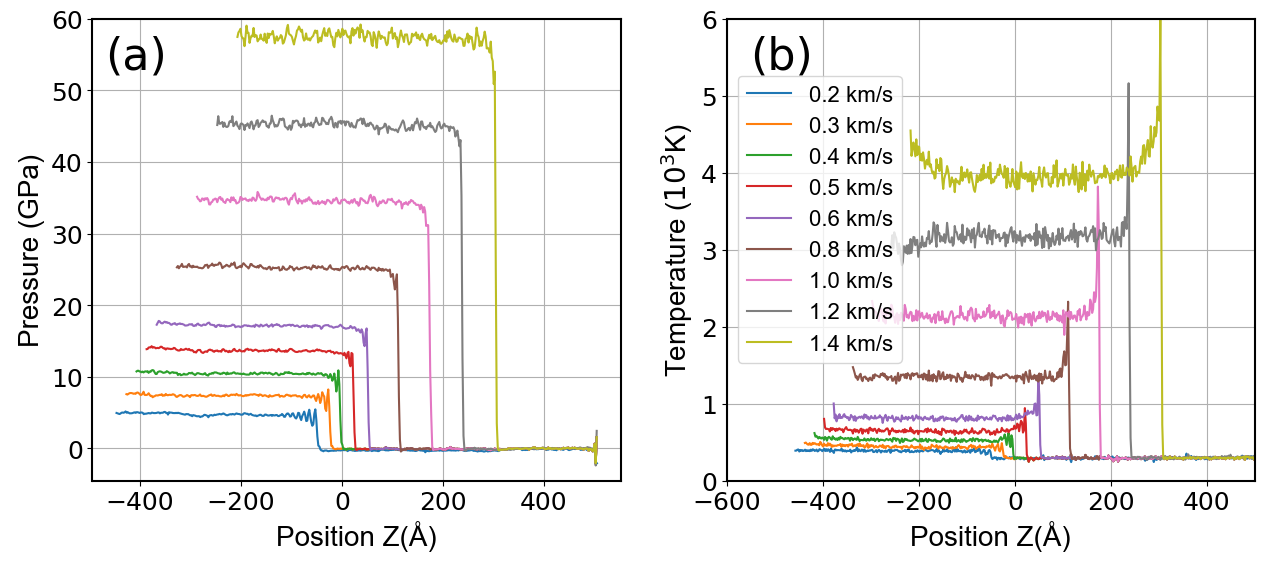}\\[8pt]
\parbox[c]{15.0cm}{\footnotesize{\bf Fig.~2.}   
The spatial distributions of thermodynamic quantities during loading along the [001] direction at 20~ps after piston loading. 
(a) Pressure distribution. 
(b) Local temperature distribution.}
\end{center}

\section{Results}
\subsection{Shock along [001] direction}
Figs.~2 (a) and (b) present the spatial distributions of pressure and temperature along the z-direction at a representative simulation time after shock loading along the [001] direction. 
For all simulated $v_\mathrm{p}$ values, the compressed region exhibited a relatively uniform pressure plateau behind the shock wave front. 
The temperature profiles showed peak values at the shock wave front, and gradually converged to plateau. 
The maximum (P,T) condition in our simulations was (58 GPa, 4000 K), which lay within the thermodynamic range covered by DP-PbSn.

Fig.~3 presents the distribution of shear stress and the corresponding snapshots during shock loading. The shear stress profiles effectively revealed the microstructure evolution under compression. 
At $v_\mathrm{p}=$ 0.2 km/s, Pb underwent neither plastic deformation nor solid-phase transformation, maintaining an FCC structure that exhibited characteristically high shear stress. 
With increasing $v_\mathrm{p}$, the shear stress within the compressed region was significant relaxed, accompanied by  plastic behavior. 
A notable phenomenon emerged where atoms near the shock front were transiently identified as BCC structure. 
However, these metastable BCC structures rapidly degenerated into deformed FCC phases within a few picoseconds. Stripe-like region identified as HCP structure emerged along 45° to the shock propagation direction, corresponding to stacking faults within the FCC phase. 
This transient BCC formation and subsequent reversion behavior has been previously reported by Li et al.\ucite{RN907} 

The major difference between the results of this study and those of Li et al.\ucite{RN907} was the activated deformation mechanisms in shock induced plasticity. 
In this study, only slipping was activated during [001] loading.
However, in the study by Li et al., not only slipping but also twinning was operative during plastic deformation, which was possibly caused by the unphysical negative twin formation energy (-0.51~meV/Å$^2$) predicted by the EAM potential. 
By contrast, DP-PbSn predicted nearly 1.0~meV/Å$^2$ formation energy of the twin, which was an underestimation of the DFT value 3.3~meV/Å$^2$. Twinning structure was not included in the dataset of DP-PbSn. However, even with this underestimation, the twinning was still not operative during the shock compression in our simulation.

\begin{center}
\includegraphics[width=15cm]{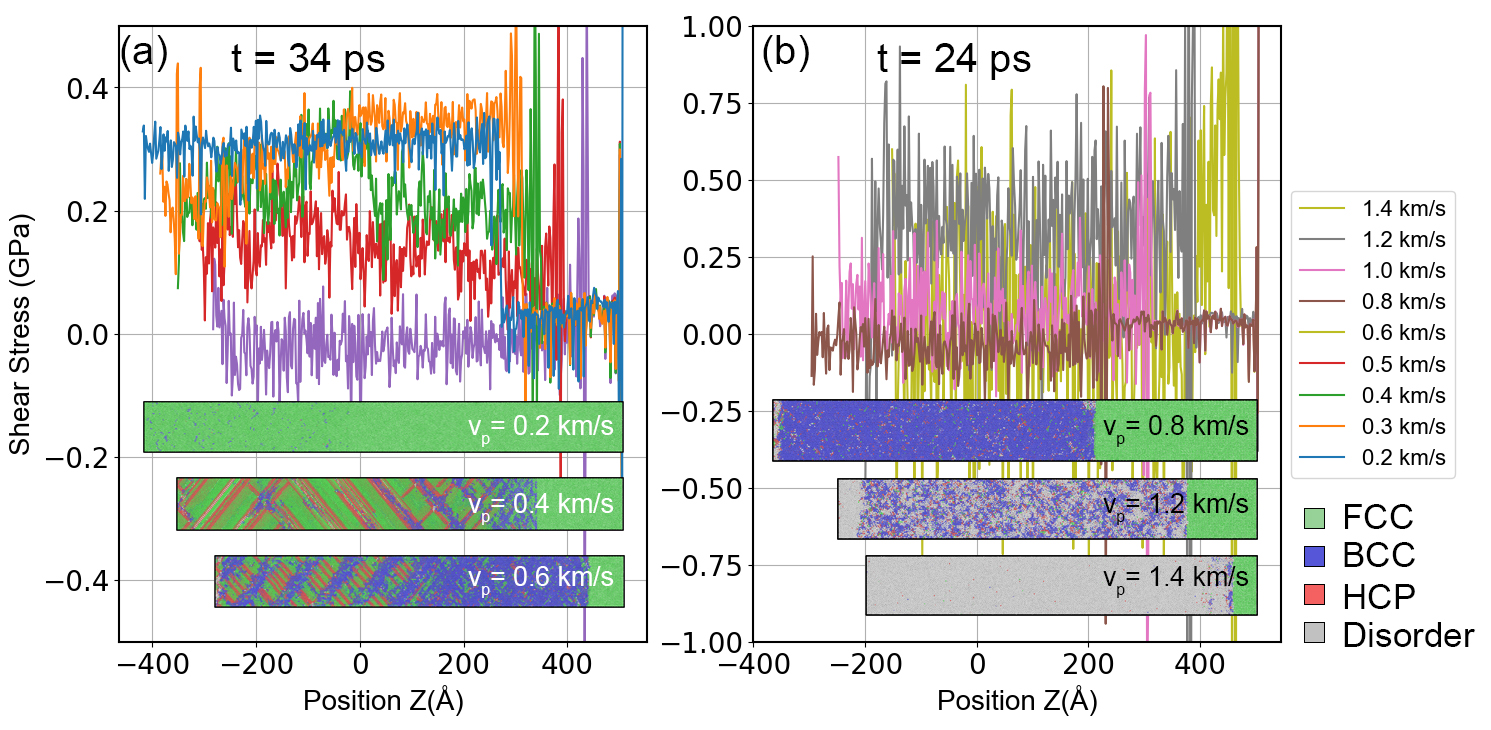}\\[8pt]
\parbox[c]{15.0cm}{\footnotesize{\bf Fig.~3.}   
The distribution of shear stress and the representative snapshots during loading along the [001] direction . 
(a) $0.2~\mathrm{km/s}\leq v_\mathrm{p} \leq 0.6~\mathrm{km/s}$ at 34 ps after piston loading. 
(b) $0.8~\mathrm{km/s}\leq v_\mathrm{p} \leq 1.4~\mathrm{km/s}$ at 24 ps after piston loading.}
\end{center}

Further increasing $v_\mathrm{p}$ drove the P-T condition in the compressed region closer to the BCC phase region on the equilibrium phase diagram. 
Consequently, the stability of shock-induced BCC structures was enhanced. 
At $v_\mathrm{p}$ = 0.6 km/s, the proportion of BCC regions within the shock wave exceeded that of deformed FCC phase, accompanied by near-zero shear stress in the compressed material. 
When $v_\mathrm{p}$ was further increased to 0.8 km/s, nearly all atoms within the compressed regions were identified BCC structures.
The reversion to FCC structures completely disappeared. 
The shear stress distribution remained approximately zero, demonstrating that both the FCC-BCC phase transformation and plastic deformation under shock loading effectively relaxed the shear stress. 
\begin{center}
\includegraphics[width=15cm]{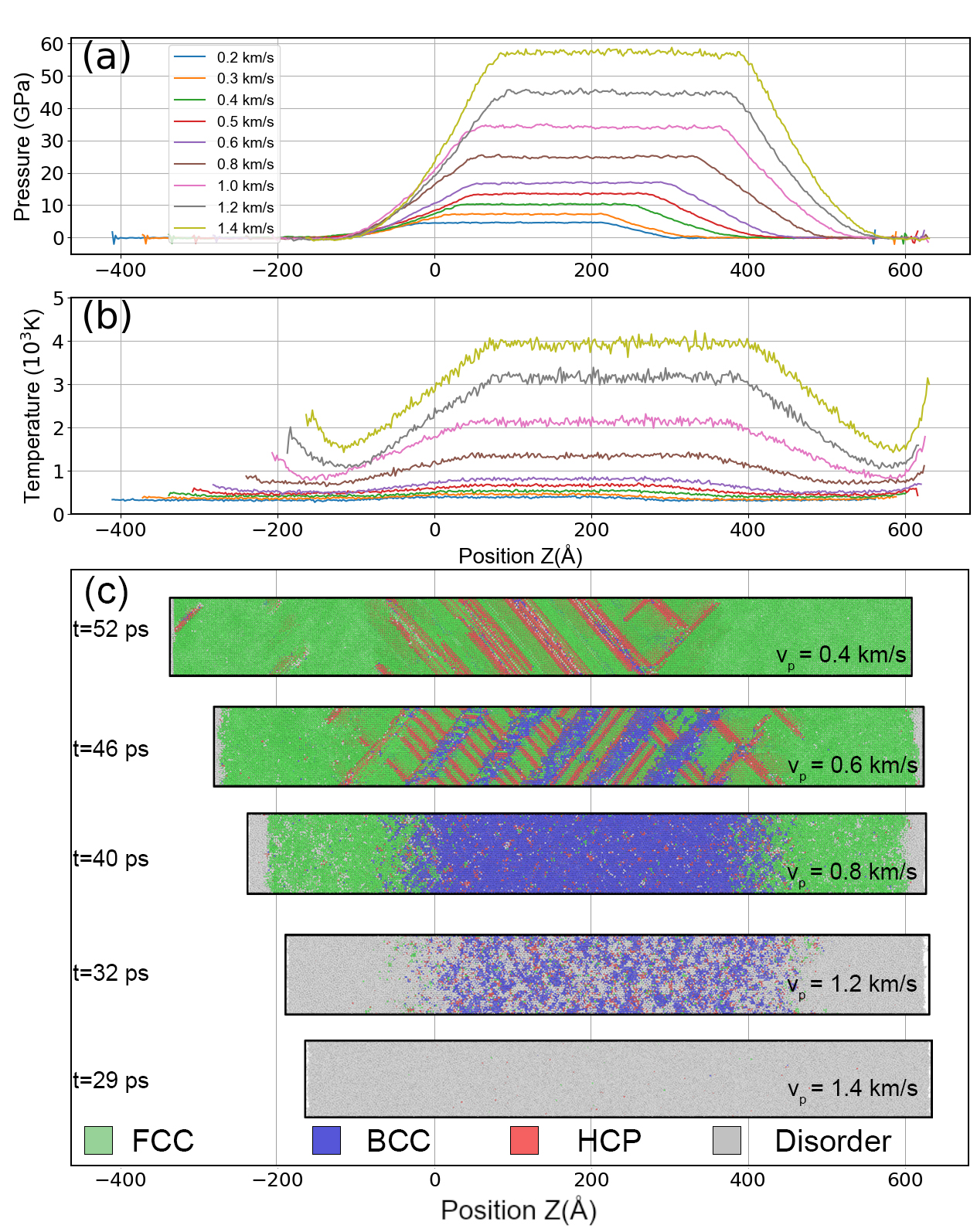}\\[8pt]
\parbox[c]{15.0cm}{\footnotesize{\bf Fig.~4.}   
(a) The distribution of pressure during [001] unloading. 
(b) The distribution of temperature during [001] unloading. 
(c) Snapshots at representative simulation times after piston loading.}
\end{center}

Interestingly, while the compressed region at $v_\mathrm{p} = 0.8~\mathrm{km/s}$ became predominantly BCC-structured, the corresponding thermodynamic conditions did not actually enter the equilibrium BCC phase region. 
This apparent contradiction stemmed from the unique FCC-BCC transition pathway under shock loading along the [001] direction. 
The Bain-path\ucite{alippi1997strained} for the FCC-BCC transformation required substantial compression along the [001] direction, which experienced a tremendously high energy barrier for the near-equilibrium processes. 
However, in dynamic processes, shock compression along the [001] orientation drove FCC deformation that aligned with the Bain path, enabling a forced transformation from FCC to BCC within picoseconds. 
Once formed, BCC phase was maintained if it was dynamically stable in the reached (P,T) condition, or reverted to the deformed FCC structure if the (P,T) was too low.

Further increasing $v_\mathrm{p}$ led to a gradually diminished shock-induced BCC phase. 
As shown in Fig.~3 (b), at $v_\mathrm{p}$ = 1.2 km/s, atoms in the compressed regions became increasingly disordered. 
When $v_\mathrm{p}$ reached 1.4 km/s, almost all atoms within the shock front became disordered. 
The corresponding thermodynamic condition (58~GPa, 4000~K, as shown in Fig. 2) exceeded the melting line in the equilibrium phase diagram\ucite{RN913}, indicating the shock-induced melting.

The above processes showed a typical phase transformation from FCC to BCC and shock-induced melting. 
No FCC to HCP phase transformation was observed. 
It is well established that this phase transformation is sluggish even at experimental timescales. 
Therefore, we considered that within the short timescales accessible in NEMD simulations, substantial nucleation of the HCP phase was unlikely to be observed. 
Moreover, the FCC-to-HCP phase transformation is known to be accomplished by the sequential glide of Shockley partial dislocations on every other close-packed \{111\} plane in the FCC lattice.\ucite{WAITZ1997837}
This mechanism requires a higher dislocation density than that observed in our simulation results, which explains the absence of nucleation in our study. 
This discrepancy was likely attributable to the use of a perfect single crystal in the molecular dynamics simulation. 
In real materials, the presence of defects across various scales can act as dislocation sources under rapid deformation, potentially generating a higher density of dislocations and thus initiating HCP nucleation.

Upon reaching the free surface, the shock wave generated a rarefaction wave that propagating in the opposite direction. 
At the same time, the piston was unloaded, leading to the creation of a release wave. 
Fig.~4 illustrates the distribution of pressure and temperature alongside with the typical snapshots during the unloading process. 
Following the rarefaction wave and the release wave, temperature and pressure in the unloaded region gradually decreased.
At $v_\mathrm{p}\leq0.8~\mathrm{km/s}$, the unloaded region predominantly retained an FCC structure. 
Stacking faults and the BCC phase in most of the unloaded region disappeared. 
The reversion from BCC to FCC during unloading followed the reverse Bain path, wherein the BCC lattice underwent significant uniaxial tensile deformation along its [001] direction to form the FCC phase. 
At $ v_\mathrm{p} \geq 1.0~\mathrm{km/s}$, the unloaded region exhibited a disordered phase rather than the FCC phase, demonstrating characteristic features of release melting.

\begin{center}
\includegraphics[width=15cm]{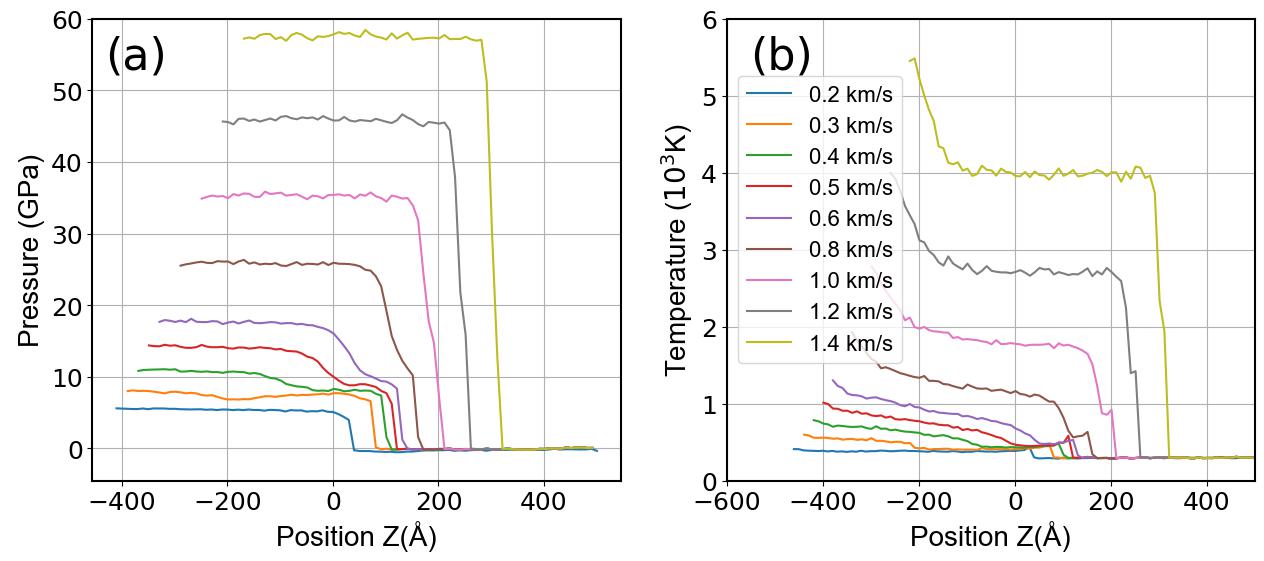}\\[8pt]
\parbox[c]{15.0cm}{\footnotesize{\bf Fig.~5.}   
The spatial distributions of thermodynamic quantities during loading along the [011] direction at 20~ps after piston loading. 
(a) Pressure distribution. 
(b) Local temperature distribution.}
\end{center}

\subsection{Shock along [011] direction}
Shock loading along the [001] direction induces an FCC-BCC transition with deformation consistent with the Bain path. 
When this consistency is not satisfied, FCC-BCC transition behavior is expected to be different.  
Moreover, when loading along other orientations, the activated slip plane may not lie perfectly 45° to the shock propagation direction, where the shear stress is maximum.
Thus, anisotropy in Pb's shock-induced plasticity and phase transition is expected. 
To further investigate this anisotropy, additional simulations were conducted with the piston loading along the [011] orientation.

Fig.~5 shows the pressure and temperature distributions during shock loading along the [011] direction. 
Compared with the loading along the [001] orientation, the compressed region along the [011] loading exhibits a more complex thermodynamic state evolution.

In the velocity range of $0.3~\mathrm{km/s} \leq v_\mathrm{p} \leq 1.0~\mathrm{km/s}$, the pressure in the compressed region underwent a relaxation period before reaching its plateau value. 
In comparison to the [001] loading cases, these relaxations were significantly slower.
The temperature within the compressed region exhibited a distinct gradient, indicating that the compressed material did not immediately achieve a Hugoniot equilibrium state upon compression, but underwent complex evolution. 
At $v_\mathrm{p}\geq 1.2~\mathrm{km/s}$, the evolution of (P,T) distribution reproduced its behavior under the [001] loading case. 
The pressure and temperature reached their plateaus immediately after the compression.

\begin{center}
\includegraphics[width=15cm]{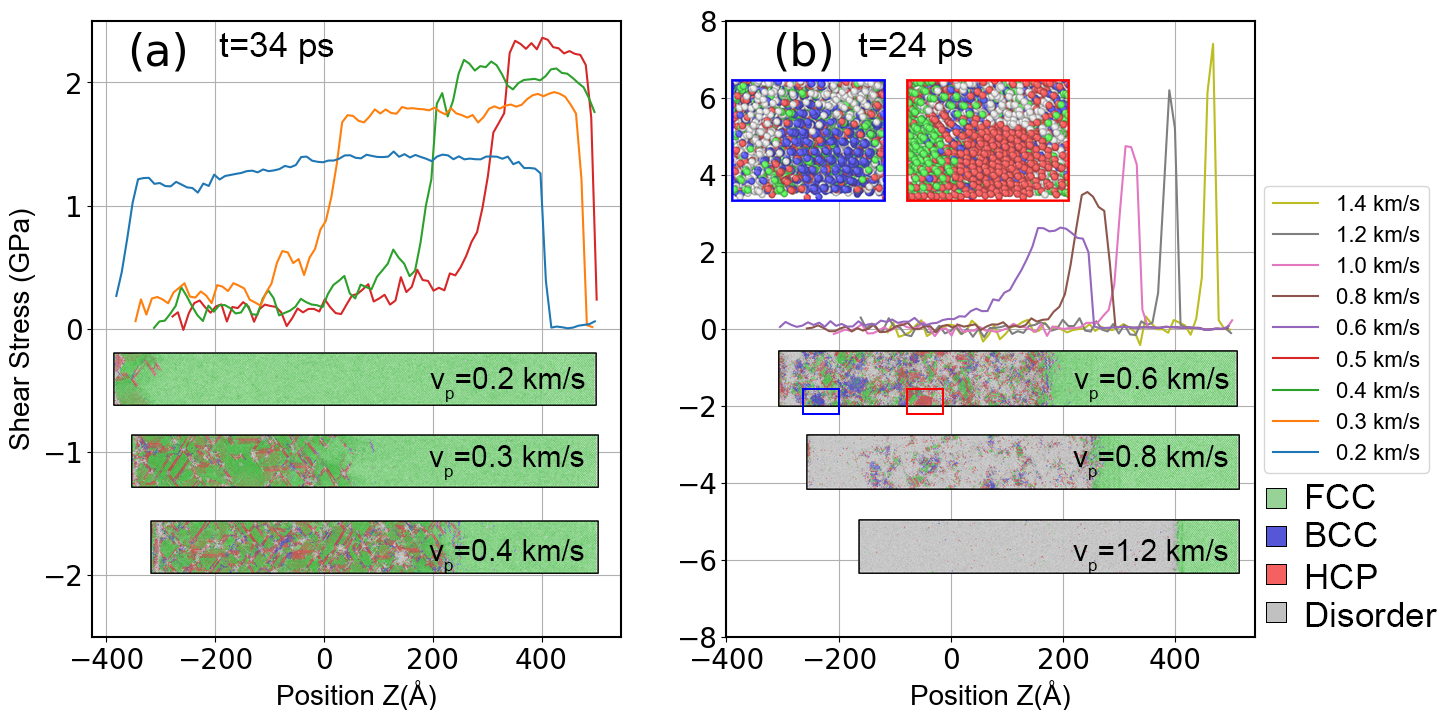}\\[8pt]
\parbox[c]{15.0cm}{\footnotesize{\bf Fig.~6.}   
The distributions of shear stress and the representative snapshots during loading along the [011] direction at 20~ps after piston loading. 
(a) $0.2~\mathrm{km/s}\leq v_\mathrm{p} \leq 0.4~\mathrm{km/s}$ at 34~ps after piston loading. 
(b) $0.6~\mathrm{km/s}\leq v_\mathrm{p} \leq 1.4~\mathrm{km/s}$ at 24~ps after piston loading. The blue and red rectangles mark the regions where the bulk BCC and HCP phases are formed.}
\end{center}

The shear stress distributions as well as the representative snapshots of NEMD simulations in the [011] loading cases at various $v_\mathrm{p}$ are shown in Fig.~6.
At $0.2~\mathrm{km/s}\leq v_\mathrm{p}\leq0.4~\mathrm{km/s}$ (Fig.~6 (a)),  stacking fault structures emerged along 60° orientations to the shock wave propagation direction. 
When $v_\mathrm{p}$ was further increased to $0.6~\mathrm{km/s}$(Fig.~6 (b)), the stacking fault structure became less prevalent, instead, bulk HCP-phase (the red rectangles in Fig.~6 (b)) domains as well as the bulk BCC-phase (the blue rectangles in Fig.~6 (b)) appeared in the compressed region.
Meanwhile, the proportion of disordered structures showed a marked increase.
With a further increase in $v_\mathrm{p}$, the proportion of disordered structures within the shock wave continued to increase. 
When $v_\mathrm{p} \geq 1.2~\mathrm{km/s}$, the compressed region became almost entirely disordered. 
Meanwhile, the shear stress became nearly zero in the compressed region, showing clear signatures of shock induced melting.

\begin{center}
\includegraphics[width=15cm]{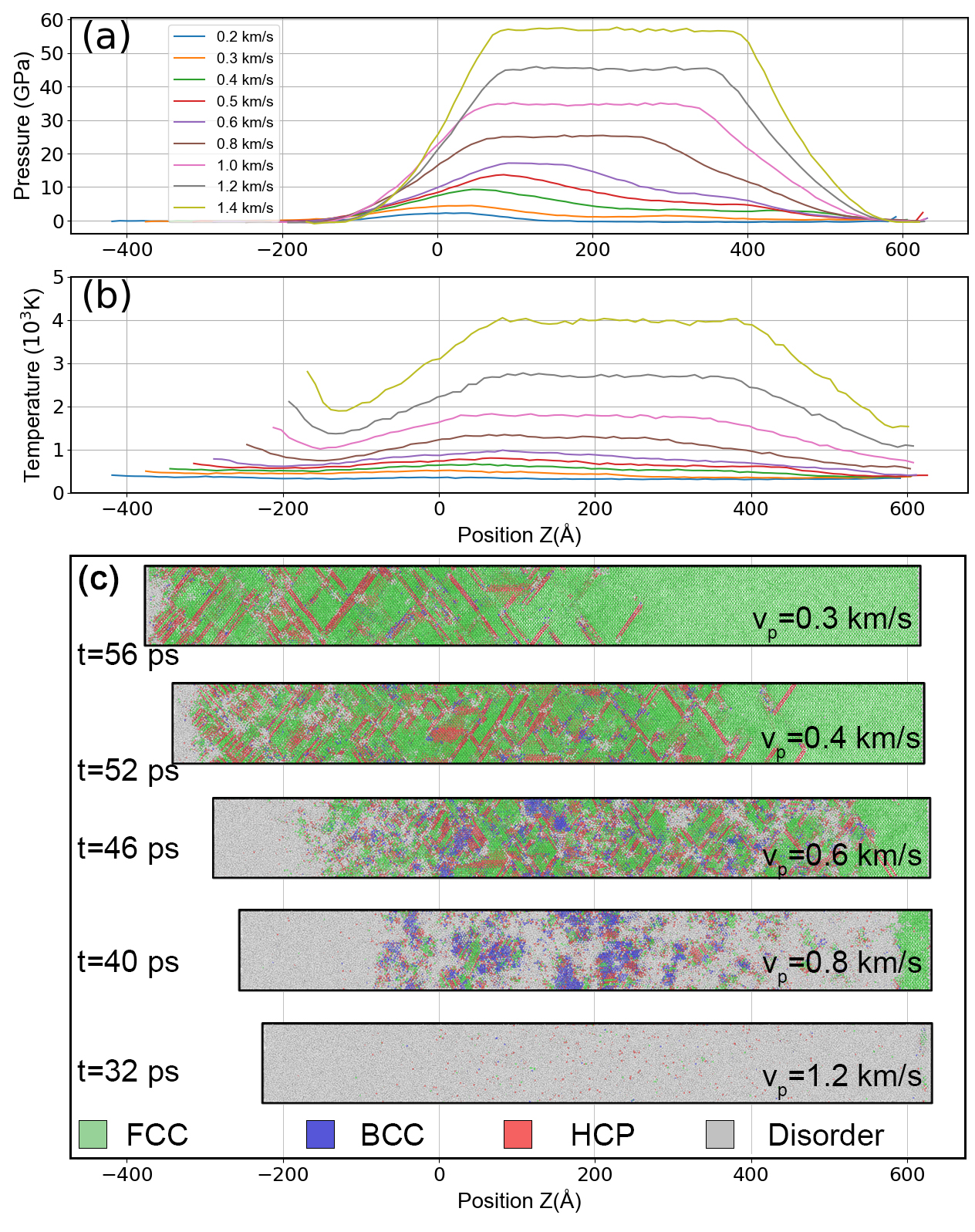}\\[8pt]
\parbox[c]{15.0cm}{\footnotesize{\bf Fig.~7.}   
(a) Pressure distribution during [011] unloading. 
(b) Temperature distribution during [001] unloading. 
(c) Snapshots at representative simulation times after piston loading.}
\end{center}

\begin{center}
\includegraphics[width=15cm]{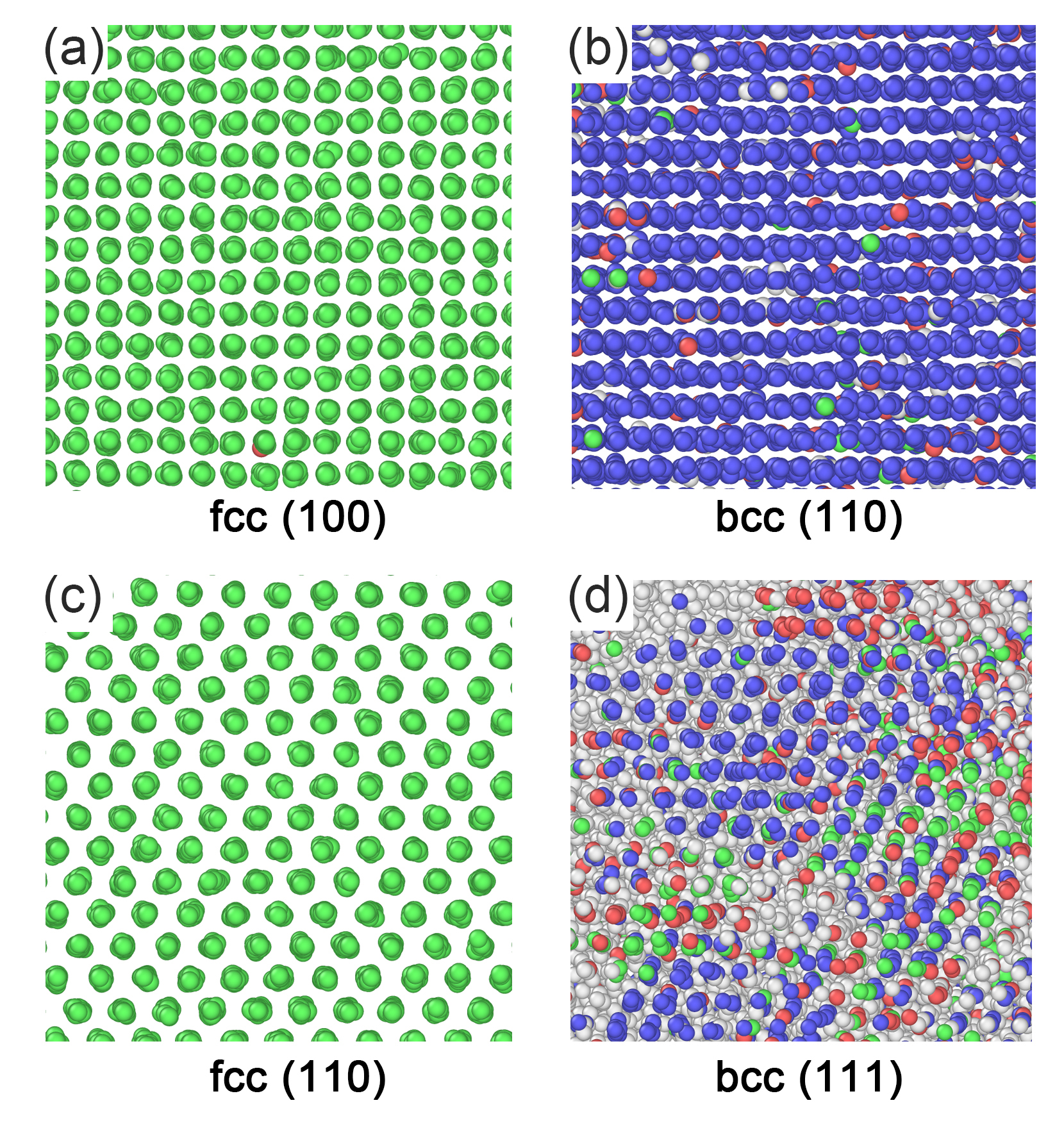}\\[8pt]
\parbox[c]{15.0cm}{\footnotesize{\bf Fig.~8.}   
Orientation relationship during dynamic loading. 
In [001] loading, the (100) plane along the initial FCC phase shown in (a) is aligned along with the (110) plane in the BCC phase in the compressed region shown in (b). 
In [011] loading, the (110) plane along the initial FCC phase shown in (c) is aligned along with the (111) plane in the BCC phase in the compressed region shown in (d).}
\end{center}

The pressure and temperature distributions, together with representative snapshots during the unloading along the [011] direction are shown in Fig.~7.
Pb exhibits significantly different behavior in [011] unloading compared to the [001] unloading. 
The key difference lies in the distribution of stacking faults. 
As shown in Figs.~7 (a) and (b), while the temperature and pressure gradually decay in the released region, the stacking faults generated during loading stage were not instantly eliminated upon unloading. 
For example, as shown in Fig.~7 (c), the unloaded region at $v_\mathrm{p}=0.3, 0.4$ and 0.6 $\mathrm{km/s}$ retains a substantial density of such plastic deformation defects identified as layered HCP structures, though the corresponding pressure had already decayed to zero.
At $v_\mathrm{p}\geq0.8~\mathrm{km/s}$ atoms in the unloaded region were identified as the disordered structure, showing significant characteristics of release melting.

The anisotropy of plastic deformation can be explained by the variations of Schmidt factor of activated slip systems. 
In [001] loading the activated slip systems are $[011]/(1-11)$ and $[0\bar11]/(111)$.
Under the uniaxial compression and tension, these slip systems exhibit the maximum Schmidt factor 0.5.
This indicates that this slip system can be subjected to the maximum resolved shear stress from the applied stress, which facilitates the activation of the slip systems.
However, in [011] loading, the two primarily activated slip systems are $[110]/(1-11)$ and $[\bar110]/(111)$. The Schmidt factor of these slip systems are 0.433, which causes a reduction of the resolved shear stresses on these slip systems.

The anisotropy of shock-induced FCC-BCC phase transformation manifests not only in the proportion of phase transition region, but also in the crystallographic orientation relationships. 
The FCC and BCC phases in the region under the [001] shock loading are shown in Figs.~8 (a) and (b), respectively, exhibiting the orientation relationship: $[001]_\mathrm{FCC}||[001]_\mathrm{BCC}$; $[100]_\mathrm{FCC}||[110]_\mathrm{BCC}$. 
This orientation relationship is typical of the well-known Bain-Path.
However, as shown in Figs.~8 (c) and (d), the FCC and BCC phases under [011] loading exhibit a distinct orientation relationship approximated by $[110]_\mathrm{FCC}||[111]_\mathrm{BCC}$ and $[100]_\mathrm{FCC}||[110]_\mathrm{BCC}$, which is in agreement with the orientation relationship discovered by Pitsch.\ucite{pitsch1959martensite} 
Nevertheless, the NEMD simulations do not provide further insight into the detailed transformation pathways of these processes.

\section{Conclusion}

To elucidate the atomistic scale processes governing the microstructure evolution of Pb in response to shock loading, we employed a high-accuracy MLP for Pb-Sn alloys to investigate shock-induced plasticity, phase transformations, melting, and defect evolution during loading-unloading processes under shock impact along both [001] and [011] directions. 
The principal findings are as follows:

1. Shock plasticity exhibits pronounced anisotropy. Under [001] loading, rapid stacking fault formation occurs, which is subsequently eliminated during unloading. 
In contrast, [011] loading generates stacking faults more slowly, yet most of these defects persist upon unloading. 
This anisotropy arises from the significantly larger Schmidt factors on operative slip systems during [001] loading.

2. FCC-BCC phase transformation during shock loading also exhibits strong anisotropy. 
Under [001] loading, the transformation follows the Bain path and reverts to the FCC phase during unloading. 
In contrast, [011] directional loading leads to the FCC-BCC phase transition following the Pitsch orientation relationship. 

The findings presented in this study establish a crucial theoretical foundation for subsequent dynamic experiments investigating lead's shock-induced phase behavior. 
The techniques used in this study can be applied to the study of the dynamic processes of other materials of interest in this field, such as Al, Cu and Fe. 
Our future work will systematically investigate the transformation pathway and underlying mechanisms associated with this unique orientation relationship.

\section*{Acknowledgment}
The work was supported by the National Key R $\&$ D Program of China under Grant No. 2022YFA1004300, 
and the National Natural Science Foundation of China under Grant No. 12404004. 

\bibliographystyle{iopart-num}
\bibliography{ref}

@article{takahashi1969lead,
  title={Lead: X-ray diffraction study of a high-pressure polymorph},
  author={Takahashi, Taro and Mao, Ho Kwang and Bassett, WA},
  journal={Science},
  volume={165},
  number={3900},
  pages={1352},
  year={1969},
  publisher={American Association for the Advancement of Science},
  doi={10.1126/science.165.3900.1352},
}

@article{KUZNETSOV2002125,
  title={FCC--HCP phase boundary in lead},
  author={Kuznetsov, A and Dmitriev, V and Dubrovinsky, L and Prakapenka, V and Weber, H-P},
  journal={Solid State Commun.},
  volume={122},
  number={3-4},
  pages={125},
  year={2002},
  doi = {https://doi.org/10.1016/S0038-1098(02)00112-6},
  publisher={Elsevier}
}

@article{RN917,
   author = {Sharma, Surinder M. and Turneaure, Stefan J. and Winey, J. M and Gupta, Y. M},
   title = {What Determines the fcc-bcc Structural Transformation in Shock Compressed Noble Metals?},
   journal = {Phys. Rev. Lett.},
   volume = {124},
   number = {23},
   pages = {235701},
   doi = {10.1103/PhysRevLett.124.235701},
   year = {2020},
   publisher = {American Physical Society},
   type = {Journal Article}
}

@article{RN913,
   author = {Krygier, A. and Powell, P. D and McNaney, J. M and Huntington, C. M and Prisbrey, S. T and Remington, B. A and Rudd, R. E and Swift, D. C and Wehrenberg, C. E and Arsenlis, A. and Park, H. S. and Graham, P. and Gumbrell, E. and Hill, M. P and Comley, A. J and Rothman, S. D},
   title = {Extreme Hardening of Pb at High Pressure and Strain Rate},
   journal = {Phys. Rev. Lett.},
   volume = {123},
   number = {20},
   pages = {205701},
   doi = {10.1103/PhysRevLett.123.205701},
   year = {2019},
   type = {Journal Article}
}

@article{RN919,
   author = {Chen, Y. and Ren, G. and Tang, T. and Li, Q. and Hu, H.},
   title = {Experimental study of micro-spalling fragmentation from melted lead},
   journal = {Shock Waves},
   volume = {26},
   number = {2},
   pages = {221},
   doi = {10.1007/s00193-015-0601-4},
   year = {2016},
   type = {Journal Article}
}

@article{RN921,
   author = {Antipov, M. V. and Arinin, V. A. and Georgievskaya, A. B. and Gnutov, I. S. and Zamyslov, D. N. and Kalashnikov, D. A. and Lebedeva, M. O. and Lebedev, A. I. and Mikhailov, A. L. and Ogorodnikov, V. A. and Panov, K. N. and Pupkov, A. S. and Rayevskiy, V. A. and Sokolova, A. S. and Syrunin, M. A. and Tkachenko, B. I. and Utenkov, A. A. and Fedorov, A. V. and Finyshin, S. A. and Chudakov, E. A. and Yurtov, I. V.},
   title = {Experimental and Computational Damage and Ejecta Studies of Pb Explosively Shock Loaded to $$P_{SL} \approx 32$$- to 40-GPa},
   journal = {J. Dyn. Behav. Mater.},
   volume = {3},
   number = {2},
   pages = {300},
   doi = {10.1007/s40870-017-0113-7},
   year = {2017},
   type = {Journal Article}
}

@article{RN909,
   author = {Xiang, Meizhen and Hu, Haibo and Chen, Jun},
   title = {Spalling and melting in nanocrystalline Pb under shock loading: Molecular dynamics studies},
   journal = {J. Appl. Phys.},
   volume = {113},
   number = {14},
   pages = {144312},
   DOI = {10.1063/1.4799388},
   year = {2013},
   type = {Journal Article}
}

@article{RN910,
   author = {Xiang, Meizhen and Hu, Haibo and Chen, Jun},
   title = {Molecular Dynamics Studies of Spalling and Melting in Shocked Nanocrystalline Pb},
   journal = {Key Eng. Mater.},
   volume = {577-578},
   pages = {613},
   DOI = {10.4028/www.scientific.net/KEM.577-578.613},
   year = {2013},
   type = {Journal Article}
}

@article{RN911,
   author = {Xiang, Meizhen and Hu, Haibo and Chen, Jun and Long, Yao},
   title = {Molecular dynamics simulations of micro-spallation of single crystal lead},
   journal = {Model. Simul. Mater. Sci. Eng.},
   volume = {21},
   pages = {055005},
   number = {5},
   DOI = {10.1088/0965-0393/21/5/055005},
   year = {2013},
   type = {Journal Article}
}

@article{wang2019atomic,
  title={An atomic view on spall responses of release melted lead induced by decaying shock loading},
  author={Wang, Kun and Zhang, Fengguo and He, Anmin and Wang, Pei},
  journal={J. Appl. Phys.},
  volume={125},
  number={15},
  pages={155107},
  year={2019},
  doi={10.1063/1.5081920},
  publisher={AIP Publishing}
}

@article{Mayer2020,
   author = {Mayer, Alexander E. and Mayer, Polina N.},
   title = {Strain rate dependence of spall strength for solid and molten lead and tin},
   journal = {Int. J. Fract.},
   volume = {222},
   number = {1-2},
   pages = {171},
   DOI = {10.1007/s10704-020-00440-8},
   year = {2020},
   type = {Journal Article}
}

@article{RN907,
   author = {Li, Guomeng and Wang, Yabin and Wang, Kun and Xiang, Meizhen and Chen, Jun},
   title = {Shock induced plasticity and phase transition in single crystal lead by molecular dynamics simulations},
   journal = {J. Appl. Phys.},
   volume = {126},
   number = {7},
   pages = {075902},
   DOI = {10.1063/1.5097621},
   year = {2019},
   type = {Journal Article}
}

@article{RN912,
   author = {Wang, Kun and Zhu, Wenjun and Xiang, Meizhen and Xu, Yun and Li, Guomeng and Chen, Jun},
   title = {Improved embedded-atom model potentials of Pb at high pressure: application to investigations of plasticity and phase transition under extreme conditions},
   journal = {Model. Simul. Mater. Sci. Eng.},
   volume = {27},
   number = {1},
   pages = {015001},
   year = {2018},
   doi = {10.1088/1361-651X/aaea55},
   type = {Journal Article}
}

@article{PhysRevB.43.1795,
  title = {Theory of high-pressure phases of Pb},
  author = {Liu, Amy Y. and Garca, Alberto and Cohen, Marvin L. and Godwal, B. K. and Jeanloz, Raymond},
  journal = {Phys. Rev. B},
  volume = {43},
  pages = {1795},
  numpages = {0},
  year = {1991},
  month = {Jan},
  publisher = {American Physical Society},
  doi = {10.1103/PhysRevB.43.1795},
}

@article{cui2008first,
  title={First-principles study of phase transition of tin and lead under high pressure},
  author={Cui, Shouxin and Cai, Lingcang and Feng, Wenxia and Hu, Haiquan and Wang, Changzheng and Wang, Yuanxu},
  journal={Phys Status Solidi B},
  volume={245},
  number={1},
  pages={53},
  year={2008},
  doi={10.1002/pssb.200743240},
  publisher={Wiley Online Library}
}

@article{PhysRevB.73.140103,
  title = {High-pressure melting of lead},
  author = {Cricchio, F. and Belonoshko, A. B. and Burakovsky, L. and Preston, D. L. and Ahuja, R.},
  journal = {Phys. Rev. B},
  volume = {73},
  pages = {140103},
  numpages = {4},
  year = {2006},
  month = {Apr},
  publisher = {American Physical Society},
  doi = {10.1103/PhysRevB.73.140103},
}

@article{vohra1990static,
  title={Static compression of metals Mo, Pb, and Pt to 272 GPa: Comparison with shock data},
  author={Vohra, Yogesh K and Ruoff, Arthur L},
  journal={Phys. Rev. B},
  volume={42},
  number={13},
  pages={8651},
  year={1990},
  publisher={APS}
}

@article{PhysRevB.70.155417,
  title = {First-principles study of low index surfaces of lead},
  author = {Yu, Dengke and Scheffler, Matthias},
  journal = {Phys. Rev. B},
  volume = {70},
  pages = {155417},
  numpages = {8},
  year = {2004},
  month = {Oct},
  publisher = {American Physical Society},
  doi = {10.1103/PhysRevB.70.155417},
}

@article{bombis2002absolute,
  title={Absolute surface free energies of Pb},
  author={Bombis, Ch and Emundts, A and Nowicki, M and Bonzel, HP},
  journal={Surf. Sci.},
  volume={511},
  number={1-3},
  pages={83},
  year={2002},
  publisher={Elsevier}
}

@article{thompson2015spectral,
	title        = {Spectral neighbor analysis method for automated generation of quantum-accurate interatomic potentials},
	author       = {Thompson, Aidan P and Swiler, Laura P and Trott, Christian R and Foiles, Stephen M and Tucker, Garritt J},
	year         = 2015,
	journal      = {J. Comput. Phys.},
	publisher    = {Elsevier},
	volume       = {285},
	pages        = {316}
}

@article{shapeev2016moment,
	title        = {Moment tensor potentials: A class of systematically improvable interatomic potentials},
	author       = {Shapeev, Alexander V},
	year         = 2016,
	journal      = {MULTISCALE MODEL SIM},
	publisher    = {SIAM},
	volume       = 14,
	number       = 3,
	pages        = {1153}
}

@article{behler2007generalized,
	title        = {Generalized neural-network representation of high-dimensional potential-energy surfaces},
	author       = {Behler, J{\"o}rg and Parrinello, Michele},
	year         = 2007,
	journal      = {Phys. Rev. Lett.},
	publisher    = {APS},
	volume       = 98,
	number       = 14,
	pages        = 146401
}

@article{bartok2010gaussian,
	title        = {Gaussian approximation potentials: The accuracy of quantum mechanics, without the electrons},
	author       = {Bart{\'o}k, Albert P and Payne, Mike C and Kondor, Risi and Cs{\'a}nyi, G{\'a}bor},
	year         = 2010,
	journal      = {Phys. Rev. Lett.},
	publisher    = {APS},
	volume       = 104,
	number       = 13,
	pages        = 136403
}

@article{chmiela2017machine,
	title        = {Machine learning of accurate energy-conserving molecular force fields},
	author       = {Chmiela, Stefan and Tkatchenko, Alexandre and Sauceda, Huziel E and Poltavsky, Igor and Sch{\"u}tt, Kristof T and M{\"u}ller, Klaus-Robert},
	year         = 2017,
	journal      = {Sci. Adv.},
	publisher    = {American Association for the Advancement of Science},
	volume       = 3,
	number       = 5,
	pages        = {e1603015}
}

@inproceedings{schutt2017schnet,
  author       = {Sch{\"u}tt, K. and Kindermans, P. J. and Sauceda, H. E. and Chmiela, S. and Tkatchenko, A. and M{\"u}ller, K. R.},
  booktitle    = {Advances in Neural Information Processing Systems 30, December 4--9, 2017, Long Beach, CA, USA},
  year         = {2017},
  pages        = {992},
}

@article{smith2017ani,
	title        = {{{ANI-1}: an extensible neural network potential with DFT accuracy at force field computational cost}},
	author       = {Smith, Justin S and Isayev, Olexandr and Roitberg, Adrian E},
	year         = 2017,
	journal      = {Chem. Sci.},
	publisher    = {Royal Society of Chemistry},
	volume       = 8,
	number       = 4,
	pages        = {3192}
}

@article{han2017deep,
	title        = {Deep Potential: a general representation of a many-body potential energy surface},
	author       = {Han, Jiequn and Zhang, Linfeng and Car, Roberto and E, Weinan},
	year         = 2018,
	journal      = {Commun. Comput. Phys.},
	volume       = 23,
	number       = 3,
	pages        = {629}
}

@inproceedings{zhang2018end,
	author       = {Zhang, Linfeng and Han, Jiequn and Wang, Han and Saidi, Wissam A and Car, Roberto and E, Weinan},
    booktitle    = {Advances in Neural Information Processing Systems 31 (NeurIPS 2018), December 3--8, 2018, Montréal, Canada},
    year         = {2018},
    pages        = {}
}

@misc{wood2017quantumaccurate,
  author       = {Mitchell A. Wood and Aidan P. Thompson},
  year         = {2017},
  note         = {arXiv:1702.07042 [physics.comp-ph]}
}

@article{gao2024neural,
  title={Neural Network Molecular Dynamics Study of Ultrafast Laser-Induced Melting of Copper Nanofilms},
  author={Gao, Tianyu and Zeng, Qiyu and Chen, Bo and Kang, Dongdong and Dai, Jiayu},
  journal={Acta Metall. Sinica},
  volume={60},
  number={10},
  pages={1439},
  year={2024}
}

@article{18-20231258,
  author       = {Zeng, Qiyu and Chen, Bo and Kang, Dongdong and Dai, Jiayu},
  title        = {大规模、量子精度的分子动力学模拟: 以极端条件液态铁为例},
  journal      = {Acta Phys. Sin.},
  volume       = {72},
  pages        = {187102-1},
  year         = {2023},
  doi          = {10.7498/aps.72.20231258},
  note = {(in Chinese)}
}

@article{liu2025machine,
  title={Machine-learning molecular dynamics simulations of shock response and spallation behavior in PPTA crystals},
  author={Liu, Lei and Shi, Jingfu and Song, Di and Miao, Changqing},
  journal={Phys. Chem. Chem. Phys.},
  volume={27},
  number={22},
  pages={11684},
  year={2025},
  publisher={Royal Society of Chemistry}
}

@article{10.1063/5.0254638,
    author = {D'Souza, J. X. and Hu, S. X. and Mihaylov, D. I. and Karasiev, V. V. and Goncharov, V. N. and Zhang, S.},
    title = {Designing a quantum-accurate machine-learning potential to enable large-scale simulations of deuterium under shock},
    journal = {Phys. Plasma},
    volume = {32},
    number = {4},
    pages = {042701},
    year = {2025},
    month = {04},
    doi = {10.1063/5.0254638},
}

@article{PhysRevB.110.224101,
  title = {Shock compression pathways to pyrite silica from machine learning simulations},
  author = {Pan, Shuning and Shi, Jiuyang and Liang, Zhixin and Liu, Cong and Wang, Junjie and Wang, Yong and Wang, Hui-Tian and Xing, Dingyu and Sun, Jian},
  journal = {Phys. Rev. B},
  volume = {110},
  pages = {224101},
  numpages = {7},
  year = {2024},
  month = {Dec},
  publisher = {American Physical Society},
  doi = {10.1103/PhysRevB.110.224101},
}

@article{Zeng2025A,
   author = {Zeng, Xin and Xiao, Shifang and Chen, Yangchun and Li, Xiaofan and Wang, Kun and Deng, Huiqiu and Zhu, Wenjun and Hu, Wangyu},
   title = {A machine-learning interatomic potential for iron under high pressure and its application to shock response},
   journal = {PHYSICA B},
   volume = {715},
   pages = {417499},
   DOI = {https://doi.org/10.1016/j.physb.2025.417499},
   year = {2025},
   type = {Journal Article}
}

@misc{DP-PbSn,
  author       = {Hou, Enze and Wang, Xiaoyang and Wang, Han},
  howpublished = {https://www.aissquare.com/models/detail?pageType=models\&name=DP-PbSn\&id=355},
  note         = {[2023-09]}
}

@article{NVNMD,
   author = {Mo, Pinghui and Li, Chang and Zhao, Dan and Zhang, Yujia and Shi, Mengchao and Li, Junhua and Liu, Jie},
   title = {Accurate and efficient molecular dynamics based on machine learning and non von Neumann architecture},
   journal = {Npj Comput. Mater.},
   volume = {8},
   number = {1},
   pages = {107},
   year = {2022},
   type = {Journal Article}
}

@article{wang2018deepmd,
	title        = {{DeePMD}-kit: A deep learning package for many-body potential energy representation and molecular dynamics},
	author       = {Han Wang and Linfeng Zhang and Jiequn Han and Weinan E},
	year         = 2018,
	journal      = {Comput. Phys. Commun.},
	publisher    = {Elsevier {BV}},
	volume       = 228,
	pages        = {178},
	doi          = {10.1016/j.cpc.2018.03.016}
}

@article{abacus1,
  title={Systematically improvable optimized atomic basis sets for ab initio calculations},
  author={Chen, Mohan and Guo, GC and He, Lixin},
  journal={J. Phys. Condens. Matter},
  volume={22},
  number={44},
  pages={445501},
  year={2010},
  publisher={IOP Publishing}
}

@article{abacus2,
  title={Large-scale ab initio simulations based on systematically improvable atomic basis},
  author={Li, Pengfei and Liu, Xiaohui and Chen, Mohan and Lin, Peize and Ren, Xinguo and Lin, Lin and Yang, Chao and He, Lixin},
  journal={Comput. Mater. Sci},
  volume={112},
  pages={503},
  year={2016},
  publisher={Elsevier}
}

@article{Perdew1996PBE,
	title        = {Generalized Gradient Approximation Made Simple},
	author       = {Perdew, John P. and Burke, Kieron and Ernzerhof, Matthias},
	year         = 1996,
	journal      = {Phys. Rev. Lett.},
	publisher    = {American Physical Society},
	volume       = 77,
	pages        = {3865}
}

@article{oncv,
  title = {Optimized norm-conserving Vanderbilt pseudopotentials},
  author = {Hamann, D. R.},
  journal = {Phys. Rev. B},
  volume = {88},
  pages = {085117},
  numpages = {10},
  year = {2013},
  month = {Aug},
  publisher = {American Physical Society},
}

@article{PD04,
title = {The PseudoDojo: Training and grading a 85 element optimized norm-conserving pseudopotential table},
journal = {Comput. Phys. Commun.},
volume = {226},
pages = {39},
year = {2018},
author = {M.J. {van Setten} and M. Giantomassi and E. Bousquet and M.J. Verstraete and D.R. Hamann and X. Gonze and G.-M. Rignanese},
}

@article{PDTh,
title = {Generating and grading 34 optimised norm-conserving Vanderbilt pseudopotentials for actinides and super-heavy elements in the PseudoDojo},
journal = {Comput. Phys. Commun.},
volume = {295},
pages = {109002},
year = {2024},
author = {Christian Tantardini and Miroslav Iliaš and Matteo Giantomassi and Alexander G. Kvashnin and Valeria Pershina and Xavier Gonze},
}

@book{marsh1980lasl,
  author       = {Marsh, Stanley P},
  title        = {LASL Shock Hugoniot Data},
  year         = {1980},
  publisher    = {University of California Press},
  address      = {Berkeley},
  volume       = {5}
}

@article{LWH_SiC,
   author = {Li, Wanghui and Hahn, Eric N. and Yao, Xiaohu and Germann, Timothy C. and Zhang, Xiaoqing},
   title = {Shock induced damage and fracture in SiC at elevated temperature and high strain rate},
   journal = {Acta Mater.},
   volume = {167},
   pages = {51},
   DOI = {10.1016/j.actamat.2018.12.035},
   year = {2019},
   type = {Journal Article}
}

@article{plimpton1995lammps,
	title        = {Fast parallel algorithms for short-range molecular dynamics},
	author       = {Plimpton, Steve},
	year         = 1995,
	journal      = {J. Comput. Phys.},
	publisher    = {Elsevier},
	volume       = 117,
	number       = 1,
	pages        = {1}
}

@article{Larsen_2016,
doi = {10.1088/0965-0393/24/5/055007},
year = {2016},
month = {may},
publisher = {IOP Publishing},
volume = {24},
number = {5},
pages = {055007},
author = {Larsen, Peter Mahler and Schmidt, Søren and Schiøtz, Jakob},
title = {Robust structural identification via polyhedral template matching},
journal = {Model. Simul. Mater. Sci. Eng.},
}

@article{ovito,
  title={Visualization and analysis of atomistic simulation data with OVITO--the Open Visualization Tool},
  author={Stukowski, Alexander},
  journal={Model. Simul. Mater. Sci. Eng.},
  volume={18},
  number={1},
  pages={015012},
  year={2009},
  publisher={IOP Publishing}
}

@article{alippi1997strained,
  title={Strained tetragonal states and Bain paths in metals},
  author={Alippi, Paola and Marcus, PM and Scheffler, Matthias},
  journal={Phys. Rev. Lett. },
  volume={78},
  number={20},
  pages={3892},
  year={1997},
  publisher={APS}
}

@article{WAITZ1997837,
title = {The f.c.c. to h.c.p. martensitic phase transformation in CoNi studied by TEM and AFM methods},
journal = {Acta Mater.},
volume = {45},
number = {2},
pages = {837},
year = {1997},
doi = {https://doi.org/10.1016/S1359-6454(96)00184-X},
author = {T. Waitz and H.P. Karnthaler},
}

@article{pitsch1959martensite,
  title={The martensite transformation in thin foils of iron-nitrogen alloys},
  author={Pitsch, W},
  journal={Philos. Mag.},
  volume={4},
  number={41},
  pages={577},
  year={1959},
  publisher={Taylor \& Francis}
}

@article{Kostiuchenko_2024,
doi = {10.1088/0256-307X/41/6/066101},
year = {2024},
month = {jun},
publisher = {Chinese Physical Society and IOP Publishing Ltd},
volume = {41},
number = {6},
pages = {066101},
author = {Kostiuchenko, Tatiana S. and Shapeev, Alexander V. and Novikov, Ivan S.},
title = {Interatomic Interaction Models for Magnetic Materials: Recent Advances},
journal = {Chin. Phys. Lett.},
}

@article{Zhao_2025,
doi = {10.1088/1674-1056/adc661},
year = {2025},
publisher = {Chinese Physical Society and IOP Publishing Ltd},
volume = {34},
number = {6},
pages = {066101},
author = {Zhao, Kai-Jie and Song, Zhi-Gong},
title = {General-purpose moment tensor potential for Ga–In liquid alloys towards large-scale molecular dynamics with ab initio accuracy},
journal = {Chin. Phys. B},
}

@article{Xiong_2023,
doi = {10.1088/1674-1056/ace4b4},
year = {2023},
month = {dec},
publisher = {Chinese Physical Society and IOP Publishing Ltd},
volume = {32},
number = {12},
pages = {128101},
author = {Xiong, Jia-Hao and Qi, Zi-Jun and Liang, Kang and Sun, Xiang and Sun, Zhan-Peng and Wang, Qi-Jun and Chen, Li-Wei and Wu, Gai and Shen, Wei},
title = {Molecular dynamics study of thermal conductivities of cubic diamond, lonsdaleite, and nanotwinned diamond via machine-learned potential},
journal = {Chin. Phys. B},
}

@article{zhuang2020discriminating,
  title={Discriminating high-pressure water phases using rare-event determined ionic dynamical properties},
  author={Zhuang, Lin and Ye, Qijun and Pan, Ding and Li, Xin-Zheng},
  journal={Chinese Physics Letters},
  volume={37},
  number={4},
  pages={043101},
  year={2020},
  publisher={IOP Publishing}
}

@article{chang2025h,
  title={H-He Demixing Driven by Anisotropic Hydrogen Diffusion},
  author={Chang, Xiaoju and Kang, Dongdong and Chen, Bo and Dai, Jiayu},
  journal={Chinese Physics Letters},
  year={2025}
}

@article{yang2022lattice,
  title={Lattice thermal conductivity of MgSiO3 perovskite and post-perovskite under lower mantle conditions calculated by deep potential molecular dynamics},
  author={Yang, Fenghu and Zeng, Qiyu and Chen, Bo and Kang, Dongdong and Zhang, Shen and Wu, Jianhua and Yu, Xiaoxiang and Dai, Jiayu},
  journal={Chinese Physics Letters},
  volume={39},
  number={11},
  pages={116301},
  year={2022},
  publisher={IOP Publishing}
}

@article{qiu2023anomalous,
  title={Anomalous thermal transport across the superionic transition in ice},
  author={Qiu, Rong and Zeng, Qiyu and Wang, Han and Kang, Dongdong and Yu, Xiaoxiang and Dai, Jiayu},
  journal={Chinese Physics Letters},
  volume={40},
  number={11},
  pages={116301},
  year={2023},
  publisher={IOP Publishing}
}

@article{hao2023artificial,
  title={An artificial neural network potential for uranium metal at low pressures},
  author={Hao, Maosheng and Guan, Pengfei},
  journal={Chinese Physics B},
  volume={32},
  number={9},
  pages={098401},
  year={2023},
  publisher={IOP Publishing}
}

@article{li2023theoretical,
  title={Theoretical predictions on superconducting phase above room temperature in lutetium-beryllium hydrides at high pressures},
  author={Li, Bin and Yang, Yeqian and Fan, Yuxiang and Zhu, Cong and Liu, Shengli and Shi, Zhixiang},
  journal={Chinese Physics Letters},
  volume={40},
  number={9},
  pages={097402},
  year={2023},
  publisher={IOP Publishing}
}

@article{zhang2025research,
  title={Research progress in the polymeric nitrogen with high energy density},
  author={Zhang, Jie and Chen, Guo and Zhang, Chengfeng and Xu, Yuxuan and Wang, Xianlong},
  journal={Chinese Physics Letters},
  year={2025},
  publisher={IOP Publishing}
}

\end{document}